\documentclass{article}
\usepackage{amsfonts}
\usepackage{amssymb}

\begin{document}

\title{One-dimensional Ising model with long-range and random short-range interactions}
\author{\textbf{A. P. Vieira}$^{\dagger}$\textbf{ and L. L. Gon\c{c}alves}$^{\ddagger}$\\Departamento de F\'{\i}sica da UFC, Cx. Postal 6030\\60451-970 Fortaleza (CE), Brazil\\$^{\dagger}$ {\small Present address: Instituto de F\'{\i}sica da USP,
05315-970 S\~{a}o Paulo (SP), Brazil. }\\$^{\dagger}$ {\small E-mail: apvieira@if.usp.br}\\$^{\ddagger}$ {\small E-mail: lindberg@fisica.ufc.br}}
\maketitle
\begin{abstract}
The one-dimensional Ising model in an external magnetic field with uniform
long-range interactions and random short-range interactions satisfying bimodal
annealed distributions is studied. This generalizes the random model discussed
by Paladin \textit{et al.} (J. Phys. I France 4, 1994, p. 1597). Exact results
are obtained for the thermodynamic functions at arbitrary temperatures, and
special attention is given to the induced and spontaneous magnetization. At
low temperatures the system can exist in a ``ferrimagnetic'' phase with
magnetization $0<\sigma<1$, in addition to the usual paramagnetic,
ferromagnetic and antiferromagnetic phases. For a fixed distribution of the
random variables the system presents up to three tricritical points for
different intensities of the long-range interactions. Field-temperature
diagrams can present up to four critical points.
\end{abstract}

\section{Introduction}

\label{intro}The one-dimensional Ising model with nearest-neighbor and uniform
infinite-range interactions was introduced by Nagle \cite{nagle70} and
presents interesting behaviour such as the existence of first- and
second-order paramagnetic transitions. The crossover between short- and
long-range interactions in the model was later reexamined by Kislinsky and
Yukalov \cite{kislinsky88}, and the study of the model for arbitrary
interactions was presented by Vieira and Gon\c{c}alves \cite{vieira95}. The
model is equivalent to a linear chain approximation of higher-dimensional
models, and it has been successfully applied to the study of
quasi-one-dimensional materials \cite{pires78,carvalho78}.

The presence of infinite-range interactions is responsible for the existence
of ferromagnetic ordering at finite temperatures, and is expected to induce
even richer behaviour in the presence of disorder. This is confirmed by the
analysis of the quenched site-dilute system by Slotte \cite{slotte85}, who
found that the system exhibits $11$ topologically different field-temperature
phase diagrams as $p$ and $\alpha$ are varied, where $p$ is the dilution
concentration and $\alpha$ measures the ratio between the intensities of
nearest-neighbor and infinite-range interactions. Recently, Paladin \textit{et
al.} \cite{paladin94} studied a random version of the model, at zero field, in
which nearest-neighbor interactions can assume opposite values $\pm J$ with
equal probability. In the quenched limit they showed that, as $\alpha$ is
varied, infinitely many ``ferrimagnetic'' ground state phases, with average
spontaneous magnetization $0<\sigma_{0}\le1$, are observed. Because of the
difficulties involved in the mathematical treatment of the system at arbitrary
temperatures, they also considered the annealed limit, which presents simpler
ground state behaviour, but where a ferrimagnetic phase is still present.

In a recent paper \cite{goncalves98a} the present authors considered a
generalization of the annealed model studied by Paladin \textit{et al}. by
introducing an external field and random nearest-neighbor interactions which
satisfy bimodal annealed distributions. The exact solution of the model was
presented, and results for the bond-dilute case were briefly discussed. This
paper considerably extends the results of the previous one by considering all
possible cases for the distribution of the random variables. The exact
solution of the model is reviewed in Sec. \ref{model} by using the approach of
Thorpe and Beeman \cite{thorpe76}, which maps the system onto a regular Ising
model with an effective temperature dependent interaction. The ground state
properties of the system are discussed in Sec. \ref{ground}, and it is shown
that ferrimagnetic phases do appear for a very large range of distributions.
At finite temperature the system presents a very rich behaviour, with many
phases, as discussed in Sec. \ref{finite}. Finally, in Sec. \ref{final} the
main results of the paper are summarized.

\section{The model}

\label{model}Consider a closed-chain ($N$ sites) Ising model in an external
magnetic field with uniform long-range interactions and random
nearest-neighbor (n.n.) interactions satisfying the bimodal annealed
distribution
\begin{equation}
\wp(\kappa_{j})=p\delta(\kappa_{j}-J_{A})+(1-p)\delta(\kappa_{j}-J_{B}).
\end{equation}
The Hamiltonian of the model is written as
\begin{equation}
H=-\sum_{j=1}^{N}\kappa_{j}\sigma_{j}\sigma_{j+1}-\frac{I}{N}\sum_{i,j=1}%
^{N}\sigma_{i}\sigma_{j}-h\sum_{j=1}^{N}\sigma_{j},
\end{equation}
which generalizes both the pure model \cite{nagle70,kislinsky88,vieira95} and
the random model studied by Paladin et al. \cite{paladin94}, corresponding to
$h=0$, $J_{A}=-J_{B}=J$ and $p=\frac{1}{2}$.

The exact solution of the model was given by the present authors in a previous
paper \cite{goncalves98a}, where an even more general model was considered,
for which long-range interactions satisfy a strongly correlated distribution.
Alternatively, the thermodynamic functions of the model can be easily obtained
from the general approach of Thorpe and Beeman \cite{thorpe76}. This consists
of performing the partial trace over the disorder variables, mapping the
system onto a regular Ising model with an effective n.n. interaction $K$,
whose properties are well known \cite{nagle70,kislinsky88,vieira95}. For the
random system, the free energy per spin as a function of the average
magnetization $\sigma$ is given by
$$
f_{a}(\sigma)=f(K,\sigma)-k_{B}T\left\{  p\ln\left[  \cosh(K-\beta
J_{A})-\epsilon(K)\sinh(K-\beta J_{A})\right]  \right.  +\nonumber $$
\begin{equation}
+(1-p)\ln\left[  \cosh(K-\beta J_{B})-\epsilon(K)\sinh(K-\beta
J_{B})\right] \nonumber
-\left.  \left[  p\ln p+(1-p)\ln(1-p)\right]  \right\}  ,
\end{equation}
where
\begin{equation}
f(K,\sigma)=-k_{B}T\ln\left\{  e^{K}\left[  \cosh\tilde{h}+\left(  \sinh
^{2}\tilde{h}+e^{-4K}\right)  ^{1/2}\right]  \right\}  +I\sigma^{2},
\end{equation}
with
\begin{equation}
\tilde{h}=\beta h+2\beta I\sigma\equiv\bar{h}+2\bar{I}\sigma,
\end{equation}
is the free energy of the regular model \cite{vieira95} with (uniform)
effective n.n. interaction $K$, determined from the solution of the equation
\begin{equation}
\frac{p}{\coth(K-\beta J_{A})-\epsilon(K)}+\frac{1-p}{\coth(K-\beta
J_{B})-\epsilon(K)}=0, \label{kp}%
\end{equation}
and
\begin{equation}
\epsilon(K)\equiv\left\langle \sigma_{j}\sigma_{j+1}\right\rangle _{K}%
=1-\frac{2e^{-4K}\left(  \sinh^{2}\tilde{h}+e^{-4K}\right)  ^{-1/2}}%
{\cosh\tilde{h}+\left(  \sinh^{2}\tilde{h}+e^{-4K}\right)  ^{1/2}} \label{ek}%
\end{equation}
is the nearest-neighbor spin-spin correlation function of the regular model.
As usual $T$ is the absolute temperature, $k_{B}$ is Boltzmann's constant and
$\beta^{-1}=k_{B}T$. The equilibrium value of the average magnetization
$\sigma$ is that which minimizes $f_{a}(\sigma)$, whose relative extrema are
determined from the solution of
\[
\left(  \frac{\partial f_{a}}{\partial\sigma}\right)  _{h,T}=0\Rightarrow
\]
\begin{equation}
\Rightarrow\sigma=\frac{\sinh\tilde{h}}{\left(  \sinh^{2}\tilde{h}%
+e^{-4K}\right)  ^{1/2}}. \label{sigma}%
\end{equation}
The equation of state can be calculated by determining, among the solutions of
Eq.\ref{sigma}, the absolute minimum of $f_{a}$ at given field $h$ and
temperature $T$.

Competition between short- and long-range interactions is responsible in the
pure model for the existence of first order phase transitions at both zero and
non-zero field. It is then natural to expect that it will produce even richer
behaviour in the presence of disorder. Assuming the long-range interaction to
be ferromagnetic ($I>0$) competition between interactions is obtained by
considering $J_{B}$ bonds as antiferromagnetic ($J_{B}<0$) and allowing
$J_{A}$ bonds to have arbitrary character ($J_{B}<J_{A}<+\infty$). These are
the cases considered in this work. As shown for the pure model \cite{vieira95}%
, for $I<0$ the system is fully frustrated and so no spontaneous order is
possible even at $T=0$.

\section{Ground state properties}

\label{ground}Since at $T=0$ the internal energy $E$ and the free energy are
equal, the ground state properties of the model can be obtained by looking for
those configurations which minimize $E$ for fixed values of the various
parameters. For annealed distributions the random variables can adjust
themselves so as to minimize the contribution to the free energy due to n.n.
interactions, with the sole restriction that the concentration of different
kinds of bonds be satisfied. This is in general achieved at fixed average
magnetization $\sigma$ by the formation of interaction domains \cite{thorpe76,
paladin94}, which avoids frustration effects. On the other hand, the
contributions to the internal energy due to long-range interactions $I$ and to
the external field $h$ depend only on $\sigma$ and are given by $-I\sigma^{2}$
and $-h\sigma$, respectively.

The possible situations are analyzed below.

\subsection{Case $J_{B}<J_{A}<0$}

This is the general case of antiferromagnetic n.n. interactions, due to the
symmetry of the model with respect to the transformation $p\rightarrow1-p$,
$J_{A}\leftrightarrow J_{B}$. When both long-range interactions $I$ and field
$h$ have small intensities compared to those of $J_{A}$ and $J_{B}$ the system
orders antiferromagnetically, satisfying all n.n. bonds. That structure has
its energy per site in the thermodynamic limit given by
\begin{equation}
E_{AF}=pJ_{A}+(1-p)J_{B}, \label{eaf}%
\end{equation}
where $p$ is the concentration of $J_{A}$ bonds. For intermediate intensities
of $I$ and $h$, ferromagnetic ordering may be induced in the $J_{A}$ domains,
while $J_{B}$ bonds are still strong enough to maintain local
antiferromagnetic ordering. It is easy to verify that such structure has an
average magnetization $\sigma=p$, and was called by Paladin \textit{et al.}
the ``ferrimagnetic'' structure \cite{paladin94}, whose energy is given by
\begin{equation}
E_{Ferri}=-pJ_{A}+(1-p)J_{B}-Ip^{2}-hp. \label{eferri}%
\end{equation}
For $I$ or $h$ strong enough the system orders ferromagnetically throughout.
Such structure, in which no nearest-neighbor bond is satisfied, has unit
average magnetization and energy given by
\begin{equation}
E_{Ferro}=-pJ_{A}-(1-p)J_{B}-I-h. \label{eferro}%
\end{equation}
To determine the range of parameters in which the various structures
correspond to equilibrium one has to compare the relative values of $E_{AF}$,
$E_{Ferri}$ and $E_{Ferro}$. Bearing in mind that for this kind of system
$\sigma$ and $h$ must have the same sign, one finds that the $T=0$ isotherm is
given by
\begin{equation}
\left|  \sigma\right|  =\left\{
\begin{array}
[c]{l}%
\left\{
\begin{array}
[c]{l}%
0,\mbox{ if }0<\left|  h\right|  <h_{1};\\
p,\mbox{ if }h_{1}<\left|  h\right|  <h_{2};\\
1,\mbox{ if }\left|  h\right|  >h_{2},
\end{array}
\right.  \mbox{ for }\left\{
\begin{array}
[c]{l}%
J_{B}<J_{A}<\frac{1}{2}J_{B}\mbox{ and }I<2(J_{A}-J_{B})\equiv I_{fi}\\
\mbox{or}\\
J_{A}>\frac{1}{2}J_{B}\mbox{ and }\left\{
\begin{array}
[c]{l}%
p<\frac{J_{A}}{J_{B}-J_{A}}\equiv p^{*}\mbox{ and }I<I_{fi}\\
\mbox{or}\\
p>p^{*}\mbox{ and }I<-\frac{2}{p}J_{A}%
\end{array}
\right.
\end{array}
\right.  ;\\
\\
\left\{
\begin{array}
[c]{l}%
0,\mbox{ if }0<\left|  h\right|  <h_{3};\\
1,\mbox{ if }\left|  h\right|  >h_{3},
\end{array}
\right.  \mbox{ for }\left\{
\begin{array}
[c]{l}%
J_{B}<J_{A}<\frac{1}{2}J_{B}\mbox{ or }J_{A}>\frac{1}{2}J_{B},\mbox{ }%
p<p^{*}\\
\mbox{and}\\
2(J_{A}-J_{B})<I<-2[pJ_{A}+(1-p)J_{B}]
\end{array}
\right.  ;\\
\\
\left\{
\begin{array}
[c]{l}%
p,\mbox{ if }0<\left|  h\right|  <h_{2};\\
1,\mbox{ if }\left|  h\right|  >h_{2},
\end{array}
,\right.  \mbox{ for }J_{A}>\frac{1}{2}J_{B},\mbox{ }p>p^{*}\mbox{ and }%
-\frac{2}{p}J_{A}<I<-\frac{2}{1+p}J_{B};\\
\\
1,\mbox{ if }\left|  h\right|  >0,\mbox{ for }\left\{
\begin{array}
[c]{l}%
J_{B}<J_{A}<\frac{1}{2}J_{B}\mbox{ and }I>-2[pJ_{A}+(1-p)J_{B}]\\
\mbox{or}\\
J_{A}>\frac{1}{2}J_{B}\mbox{ and }\left\{
\begin{array}
[c]{l}%
p<p^{*}\mbox{ and }I>-2[pJ_{A}+(1-p)J_{B}]\\
\mbox{or}\\
p>p^{*}\mbox{ and }I>-\frac{2}{1+p}J_{B}%
\end{array}
\right.
\end{array}
\right.  ,
\end{array}
\right.
\end{equation}
where
\begin{equation}
\left\{
\begin{array}
[c]{c}%
h_{1}=-2J_{A}-pI,\\
h_{2}=-2J_{B}-(1+p)I,\\
h_{3}=-2[pJ_{A}+(1-p)J_{B}]-I.
\end{array}
\right.
\end{equation}

\subsection{Case $J_{B}<0\leq J_{A}$}

In this case, even for $J_{A}=0$ and $I$, $h\rightarrow0^{+}$ internal energy
is minimized by a structure where antiferromagnetic ordering prevails in
$J_{B}$ domains while the rest of the system orders ferromagnetically. This
corresponds to the ferrimagnetic phase ($\sigma=p$), and so for $J_{A}\geq0$
there is no stable antiferromagnetic structure. Again for $I$ or $h$ strong
enough the system orders ferromagnetically throughout ($\sigma=1$). The
relevant internal energies per spin in the thermodynamic limit are
\begin{equation}
E_{Ferri}=-pJ_{A}+(1-p)J_{B}-Ip^{2}-hp
\end{equation}
and
\begin{equation}
E_{Ferro}=-pJ_{A}-(1-p)J_{B}-I-h,
\end{equation}
so that the $T=0$ isotherm is given by
\begin{equation}
\left|  \sigma\right|  =\left\{
\begin{array}
[c]{l}%
\left\{
\begin{array}
[c]{l}%
p\mbox{,\quad if\quad}0<|h|<-2J_{B}-(1+p)I\mbox{ }\\
1\mbox{,\quad if\quad}|h|>-2J_{B}-(1+p)I
\end{array}
,\right.  \mbox{\quad for\quad}I<-\frac{2J_{B}}{(1+p)};\\
\\
1\mbox{,\quad if\quad}|h|>0\mbox{\quad for\quad}I>-\frac{2J_{B}}{(1+p)}.
\end{array}
\right.
\end{equation}

\section{Finite temperature properties}

\label{finite}From the results of the last section, one concludes that there
are essentially three different regimes for the ground state behaviour of the
system at zero field, depending on the distribution of the random variables.
For $J_{B}<J_{A}<\frac{1}{2}J_{B}$ the spontaneous magnetization ($\sigma
_{0}\equiv\left.  \sigma\right|  _{h=0}$) can assume only the values
$\sigma_{0}=0$ and $\sigma_{0}=1$, corresponding to antiferromagnetic and
ferromagnetic phases, respectively. For $\frac{1}{2}J_{B}<J_{A}<0$
ferrimagnetic ($\sigma_{0}=p$) phases become possible, in addition to
antiferromagnetic and ferromagnetic phases. Finally, for $J_{A}\geq0$ only
ferrimagnetic and ferromagnetic phases exist in equilibrium.

The ground state properties lead to the conclusion that, at finite
temperatures, the free energy as a function of $\sigma_{0}$ presents up to
five relative minima in the interval $-1\le\sigma_{0}\le1$, corresponding to
the symmetric ferromagnetic ($\sigma_{0}\simeq\pm1$) and ferrimagnetic
($\sigma_{0}\simeq\pm p$) phases and to the antiferromagnetic (or
paramagnetic) phase ($\sigma_{0}=0$). In the presence of an external field
$h>0$ the relative minima of the free energy $f_{a}(\sigma)$ are shifted
towards the positive $\sigma$ axis and the values of $f_{a}(\sigma)$ at those
minima are altered, which may result in discontinuities or critical points in
the $\sigma\times h$ isotherms.

The second order paramagnetic transition temperature $T_{c}$ is obtained, as
in the pure model \cite{vieira95}, by taking the $\sigma\rightarrow0$ limit
with $h=0$ in Eq.\ref{sigma}, noting however that the effective interaction
$K$ depends implicitly on $\sigma$ through $\epsilon(\sigma)$. Therefore, one
obtains
\begin{equation}
e^{-2K_{0}}\equiv\left.  e^{-2K}\right|  _{\sigma=0}=2\beta_{c}I,
\end{equation}
where $\beta_{c}^{-1}=k_{B}T_{c}$. Taking the $\sigma\rightarrow0$ limit in
Eqs.\ref{kp} and \ref{ek} one also obtains
\begin{equation}
\tanh K_{0}=p\tanh(\beta_{c}J_{A})+(1-p)\tanh(\beta_{c}J_{B}), \label{tanhk0}%
\end{equation}
and by combining the two previous equations one gets
\begin{equation}
\frac{1-p\tanh(\beta_{c}J_{A})-(1-p)\tanh(\beta_{c}J_{B})}{1+p\tanh(\beta
_{c}J_{A})+(1-p)\tanh(\beta_{c}J_{B})}=2\beta_{c}I,
\end{equation}
which directly determines $T_{c}$.

As discussed for the pure model \cite{vieira95}, the tricritical point can be
found by solving the system
\begin{equation}
\left.  \frac{\partial f_{a}}{\partial\sigma}\right|  _{\sigma=\sigma_{t}%
}=0\mbox{,\quad}f_{a}(\sigma_{t})=f_{a}(0),
\end{equation}
and imposing that the $\sigma_{t}\rightarrow0$ root of the second equation be
quadruple by means of a Taylor expansion, which is equivalent to requiring
that all derivatives up to fourth order of $f_{a}$ with respect to $\sigma$ be
zero at $\sigma=0$. Taking into account the implicit dependence of $K$ on
$\sigma$, one gets
\begin{equation}
e^{-4K_{0}}-3+6K_{0}^{\prime\prime}=0, \label{ptr}%
\end{equation}
where
\begin{equation}
K_{0}^{\prime\prime}\equiv\left.  \frac{\partial^{2}K}{\partial\sigma^{2}%
}\right|  _{\sigma=0}=-8\left(  \beta I\right)  ^{2}\frac{p(1-p)\left[
\tanh\left(  \beta J_{A}\right)  -\tanh\left(  \beta J_{B}\right)  \right]
^{2}}{\left\{  1-\left[  p\tanh\left(  \beta J_{A}\right)  +(1-p)\tanh\left(
\beta J_{B}\right)  \right]  ^{2}\right\}  ^{2}}. \label{k2s}%
\end{equation}
For fixed $J_{A}$, $J_{B}$ and $p$, by solving the system composed of
Eqs.\ref{tanhk0}, \ref{ptr} and \ref{k2s} the coordinates $I_{tr}$ and
$T_{tr}$ of the tricritical point are determined. In the non-random limit
($p=0$, $p=1 $ or $J_{A}=J_{B}$) Eq.\ref{k2s} gives $K_{0}^{\prime\prime}=0$
and one obtains the pure model result \cite{nagle70, kislinsky88, vieira95}
\begin{equation}
\left\{
\begin{array}
[c]{c}%
e^{-2\beta _{tr}J}=2\beta _{tr}I_{tr}, \\
e^{-4\beta _{tr}J}-3=0.
\end{array}
\right.
\end{equation}

The following subsections analyze the behaviour of the system for different
values of $J_{A}$ and $J_{B}$, using the renormalized parameters
\begin{equation}
\delta\equiv\frac{J_{A}}{J_{B}},\mbox{\quad}\alpha\equiv\frac{I}{J_{B}}%
,\quad\gamma\equiv\frac{h}{\left|  J_{B}\right|  },\quad\theta\equiv
\frac{k_{B}T}{\left|  J_{B}\right|  }.
\end{equation}

\subsection{Case $J_{B}<J_{A}<\frac{1}{2}J_{B}<0$ ($\frac{1}{2}<\delta<1$)}

For this case and at zero field ($h=0$) the behaviour of the system is
qualitatively the same as for the pure model, regardless of the value of $p$,
since there is no stable ferrimagnetic phase. For very strong long-range
interactions $I$ the system is ferromagnetic at low temperatures and there is
a second order paramagnetic transition. As $I$ gets weaker, passing through
the tricritical point $I=I_{tr}$ paramagnetic transitions change to first
order and, for $I<I_{ferro}\equiv-2\left[  pJ_{A}+(1-p)J_{B}\right]  $ the
ground state becomes antiferromagnetic, and $\sigma_{0}=0$ for all $T$.

In the presence of a magnetic field $h>0$ and for $I<I_{fi}\equiv2(J_{A}%
-J_{B})$ the $\sigma\times h$ isotherms present two discontinuities at $T=0$
(from $\sigma=0$ to $\sigma=p$ and from $\sigma=p$ to $\sigma=1$), which
persist at low temperatures, inducing qualitative differences compared to the
pure model. Figures 1(a) and (c) show $\sigma\times\gamma$
isotherms for various values of the renormalized temperature $\theta\equiv
k_{B}T/\left|  J_{B}\right|  $ when $\delta=\frac{2}{3}$ and $p=0.5$, with
$\alpha=-0.5$ and $\alpha=-0.6$ , respectively, while corresponding
$\gamma\times\theta$ phase diagrams are shown in Figs. 1(b) and (d).
In both cases there are two discontinuities in the $T=0$ isotherm for $h>0$
(or, by symmetry, $h<0$), producing four first order lines in the phase
diagrams. For $\alpha=-0.5$ those lines remain separate, ending at critical
points corresponding to temperatures $\theta_{hc}$ and $\theta_{hc}^{\prime}$,
but for $\alpha=-0.6$ the lines can intersect, giving rise to other two
symmetric lines which terminate at a critical point. In fact, one finds that
for $\alpha_{hcr2}\simeq-0.5751$ the critical points corresponding to
$\theta_{hc}^{^{\prime}}$ touches the other first order lines, which produces
for $\alpha<\alpha_{hcr2}$ the behaviour observed in Figs. 1(c) and (d).

Figure 2 shows the $\theta\times\alpha$ diagram for $\delta
=\frac{2}{3}$ and $p=0.5$. At zero field, as already stated, the behaviour is
similar to that of the pure model, with the tricritical point separating
regions of first and second order paramagnetic transitions. At non-zero field,
on the other hand, there are two critical temperature ($\theta_{hc}$) curves
for $\alpha\rightarrow0^{-}$. One of the curves, as in the pure model, ends at
the tricritical point, giving rise to the second order transition temperature
$\theta_{c}$. The other curve ends at point $P_{hcr2} $, reaching the
$\theta_{hcr3}$ curve, which touches the $\alpha$ axis at $\alpha_{fi}%
=-\frac{2}{3}$.

\subsection{Case $\frac{1}{2}J_{B}<J_{A}<0$ ($0<\delta<\frac{1}{2}$)}

The most interesting kind of behaviour occurs for values of $\delta\in
(0,\frac{1}{2})$. The existence of ferrimagnetic phases at zero field for
$p>p^{*}\equiv J_{A}/(J_{B}-J_{A})$ is responsible for strong qualitative
changes in the phase diagrams.

For $p>p^{*}$ and $I>I_{ferri}\equiv-2J_{A}/p$ the system can undergo first
order ferri-ferromagnetic transitions as the temperature is raised, as
observed in the $\theta\times\alpha$ diagrams of Figs. 3 and
4 for $\delta=\frac{1}{3}$ and $p=0.6$ and $p=0.7$, respectively.
In both figures dotted curves show first-order ferri-ferromagnetic
temperatures $\theta_{f}$. For $p^{*}=0.5<p=0.6<p_{cr2}\simeq0.6596$ there is
coexistence of paramagnetic, ferrimagnetic and ferromagnetic phases at point
$P_{cr3}$, where the $\theta_{f}$ curve reach the first-order paramagnetic
transition temperature $\theta_{t}$. So, to the left of $P_{cr3}$ first-order
ferri-paramagnetic transition occur at $\theta_{t}$. On the contrary, for
$p=0.7>p_{cr2}$ the $\theta_{f}$ curve does not reach $\theta_{t}$, ending at
point $P_{cr}$ together with $\theta_{hc}^{\prime}$.

The behaviour of the spontaneous magnetization $\sigma_{0}$ can be understood
by analyzing the evolution of the free energy $f_{a}(\sigma_{0})$ as the
temperature is varied. For $\delta=\frac{1}{3}$ and $p=0.7$ Figs.
5(a) and (b) show $f(\sigma_{0})\equiv f_{a}(\sigma_{0})/\left|  J_{B}\right|
$ at different temperatures for $\alpha=-1.10>\alpha_{cr}\simeq-1.1018$ and
$\alpha=-1.13$, respectively. In both cases the ground state is
antiferromagnetic ($\left|  \sigma_{0}\right|  =p$) and there exist metastable
states for $\left|  \sigma_{0}\right|  =1$ and $\sigma_{0}=0$. For
$\alpha=-1.10$ the metastable ferromagnetic state disappears at $\theta
\simeq0.2$ and the spontaneous magnetization increases with temperature in a
certain range. At $\theta_{t}\simeq0.430$ the system undergoes a first order
ferri-paramagnetic transition. On the other hand, for $\alpha=-1.13$
first-order ferri-ferromagnetic and ferro-paramagnetic transitions occurs at
$\theta_{f}\simeq0.115$ and $\theta_{t}\simeq0.532$.

Zero field transitions can be better observed in Figs. 6 and
7, which show $\sigma_{0}$ as a function of the renormalized
temperature $\theta$ for the case $\delta=\frac{1}{3}$ with $p=0.6$ and
$p=0.7$, respectively. For $p=0.6$ the ground state is ferromagnetic for
$\alpha<\alpha_{ferro}=-1.25$, and the system undergoes first-order
ferro-paramagnetic transitions for $\alpha>\alpha_{tr}\simeq-1.6997$. If
$\alpha_{ferro}<\alpha<\alpha_{ferri}\simeq-1.1111$ the ground state is
ferrimagnetic, and if $\alpha<\alpha_{cr3}\simeq-1.2009$ both first-order
ferri-ferromagnetic and ferro-paramagnetic transitions occur. If
$\alpha>\alpha_{cr3}$ only first-order ferri-paramagnetic transitions are
observed. For $p=0.7$, when $\alpha_{tr}\simeq-1.4960$, $\alpha_{ferro}%
\simeq-1.1764$ and $\alpha_{ferri}\simeq-0.9523$, the behaviour is much the
same, with the important exception that ferri-ferromagnetic transition occur
if $\alpha_{ferro}<\alpha<\alpha_{cr}\simeq-1.1018$, and exactly if
$\alpha=\alpha_{cr}$ there is a point in the $\sigma_{0}\times\theta$ curve
where $\sigma_{0}$ is a continuous function, but its thermal derivative
diverges. An important feature of this model that is evident in Fig.
7 is the increase of $\sigma_{0}$ with temperature in a certain
range, which is verified in some perovskite-like materials such as Zn$_{1-x}%
$Ga$_{x}$Mn$_{3}$C ($x\le1$) and CuMn$_{3}$N \cite{heritier80}. Finally it
should be noted that for $\left|  \alpha\right|  $ greater than what is shown
in the figures the behaviour of $\sigma_{0}$ becomes identical to that of the
pure model, with the ferro-paramagnetic transition changing from first to
second order at the tricritical point.

The effects of the existence of ferrimagnetic phases at zero field on the
$\gamma\times\theta$ phase diagrams can be observed for $\delta=\frac{1}{3}$
with $p=0.6$ and $p=0.7$ in Figs. 8(a)-(f) and Figs. 9%
(b) and (d). For $p=0.6$ figures 8(a)-(f) show the evolution of
$\gamma\times\theta$ diagrams from $\alpha=-1$ to $\alpha=-1.21$. If
$\alpha>\alpha_{ferri}\simeq-1.1111$ the diagrams are similar to those shown
in Fig. 1(b). Exactly at $\alpha=\alpha_{ferri}$ there appears a
zero-field first-order line, which splits at $\theta=\theta_{f}$, giving rise
to two symmetric first-order lines. If $\alpha>\alpha_{hcr2}\simeq-1.1295$
symmetric lines intersect at $\theta=\theta_{hcr3}$, and $\gamma_{hcr3}$, the
corresponding field, becomes zero at $\alpha_{cr3}\simeq-1.2009$. For $p=0.7$
there is no intersection of symmetric first-order lines, as shown in Figs.
9(b) and (d). Figures 9(c) and (d) illustrate the
behaviour of the system if $\alpha=\alpha_{cr}$, when one of the coexistence
curves just touches the $\sigma$ axis.

As the value of $\delta$ becomes smaller a reentrant behaviour appears in the
ferri-ferromagnetic transition temperature ($\theta_{f}$) lines, as can be
seen in Fig. 10, which shows $\theta\times\alpha$ diagrams at zero
field for $\delta=\frac{1}{5}$ and various values of $p$. The reentrant
behaviour of $\theta_{f}$, evident in Figs. 10(b) and (c), is
reflected in the thermal behaviour of the spontaneous magnetization
$\sigma_{0}$ shown in Fig. 11, where two first-order
ferri-ferromagnetic transition occur for $\alpha_{cr}\simeq-1.3108<\alpha
<\alpha_{re}\simeq-1.2710$. The existence of reentrant behaviour in random
models with competing interactions is common in both annealed \cite{thorpe76} %
and quenched \cite{wolff85} systems.

Reentrant behaviour in $\theta\times\alpha$ diagrams is related to the
structure of first-order lines in $\gamma\times\theta$ phase diagrams, as
observed in Figs. 12(b) and (d), for $\delta=\frac{1}{5}$, $p=0.5$,
and $\alpha=\alpha_{re}\simeq-1.2710$ and $\alpha=-1.29$, respectively. In the
first case the symmetric upper and lower first order lines are tangent to the
$\gamma=0$ axis at $\theta=\theta_{re}$, while in the second case those lines
are composed of two distinct segments, touching the $\gamma=0$ axis at
temperatures $\theta_{f}$ and $\theta_{f}^{\prime}$, producing the
discontinuities observed in the corresponding $\sigma_{0}(\theta)$ curve in
Fig. 11.

For $0<\delta\lesssim0.1649$ the $\theta\times\alpha$ diagrams can present up
to three tricritical points for $p\in(p_{rt}^{\prime},p_{rt})$. This is
exemplified for $\delta=0.16$ and $p_{rt}^{\prime}\simeq0.4504<p=0.453<p_{rt}%
\simeq0.4567$ in Fig. 13. The existence of three tricritical points
($P_{tr}$, $P_{tr}^{\prime}$ and $P_{tr}^{\prime\prime}$) is reflected in the
appearance of another zero-field second-order paramagnetic transition line for
$\alpha_{tr}^{\prime\prime}<\alpha<\alpha_{tr}^{\prime}$. That line changes
from only one point for $p=p_{rt}^{\prime}$, where $P_{tr}^{\prime}$ and
$P_{tr}^{\prime\prime}$ coincide, until it joins the usual $\theta_{c}$ line
for $p=p_{rt}$, where $P_{tr}^{\prime\prime}$ coincides with $P_{tr}$. An
example of $\theta\times\alpha$ diagram with two tricritical points is given
in Fig. 14. It should be noted that point $P_{c2}$, which marks the
intersection of the $\theta_{f}$, $\theta_{c}^{\prime}$ and $\theta_{t}$
lines, is not a tricritical point, since all those lines correspond to $h=0$,
and so the critical exponents at $P_{c2}$ have the usual mean-field values
($\alpha=0$, $\beta=\frac{1}{2}$, $\gamma=1$, $\delta=3$, within the usual
convention of critical indices).

\subsection{Case $J_{A}\ge0$ ($\delta\le0$)}

In this situation there is no antiferromagnetic ground state, as discussed in
Sec. \ref{ground}. Irrespective of the value of $\alpha\equiv I/\left|
J_{B}\right|  $ there is always a second-order ferri- or ferro-paramagnetic
transition at a finite temperature. This is illustrated by the $\theta
\times\alpha$ diagrams of Figs. 15-17 for $\delta=0$ and
$p$ varying from $p=0.2$ to $p=0.4$, where there is at most only one
tricritical point. This happens for all $\delta\le0$, and the tricritical
point exists only for $p<p_{mc}$, where $p_{mc}\rightarrow 0$ as $\delta
\rightarrow -\infty$. Among the cases shown in the
figures, the most interesting diagram corresponds to $p=0.3$, where there are
two points $P_{c2}$ and $P_{c2}^{\prime}$ at which $\theta_{t}$, $\theta_{f}$
and $\theta_{c}$ lines intersect.

Figures 18 and 19 show $\sigma\times\gamma$ isotherms and
$\gamma\times\theta$ phase diagrams for $\delta=0$ with $p=0.2$ and $p=0.3$,
respectively. In the first case, Fig. 18(b) shows the $\gamma
\times\theta$ diagram for $\alpha=-1.5>\alpha_{re}$, and three first-order
lines are observed, one of which corresponds to $\gamma=0$. The two remaining
symmetrical lines are tangent to each other for $\alpha=\alpha_{re}$, and this
gives rise to reentrant behaviour of the spontaneous magnetization for
$\alpha<\alpha_{re}$, as previously discussed. For $\alpha=\alpha_{c2}%
\simeq-1.6549$, as shown in Figs. 18(c) and (d), the critical
temperature $\theta_{c}$ of the first-order $\gamma=0$ line coincides with
$\theta_{f}^{\prime}$, the temperature at which the symmetrical $\gamma\neq0$
lines reappear. In the case $p=0.3$, Figs. 19(b) and (d) exemplify
the behaviour of the system around point $P_{c2}^{\prime}$ (Fig.
16), which marks the reappearance of a critical (second-order)
point in the $\gamma=0$ first-order line. Such critical point is absent for
$\alpha_{c2}^{\prime}<\alpha<\alpha_{c2}$.

\section{Conclusions}

\label{final}In this paper the one-dimensional Ising model with uniform
infinite-range interactions and random nearest-neighbor interactions in an
external field was studied. The random bond $\kappa_{i}$ connecting the sites
$i$ and $i+1$ was considered to satisfy the bimodal annealed probability
distribution
\[
\wp(\kappa_{j})=p\delta(\kappa_{j}-J_{A})+(1-p)\delta(\kappa_{j}-J_{B}).
\]
Explicit results were presented for the cases where there is competition
between short- and long-range interactions. In order to avoid complete
frustration of the system long-range interaction was assumed to be
ferromagnetic. Competition was then produced by assuming $J_{B}$ to be
antiferromagnetic and letting $J_{A}$ be of arbitrary character. It was shown
that at $T=0$ the system can present antiferromagnetic, ferromagnetic and
``ferrimagnetic'' ordering. The spontaneous magnetization was shown to present
unusual behaviour such as first-order ferri-ferromagnetic transitions,
positive thermal derivative and reentrant behaviour. For certain distributions
in which $0<\delta\lesssim0.1649$, where $\delta$ is the ratio between the
possible values of antiferromagnetic nearest-neighbor interactions, the system
can present up to three tricritical points as the intensity of the
infinite-range interactions is varied. Many topologically different
field-temperature phase diagrams are possible, and these can present up to
four critical points, as opposed to the pure model which presents at most two
critical points.

\bibliographystyle{prsty}
\bibliography{mestrado}
\newpage

\noindent {\LARGE Figure captions}\bigskip

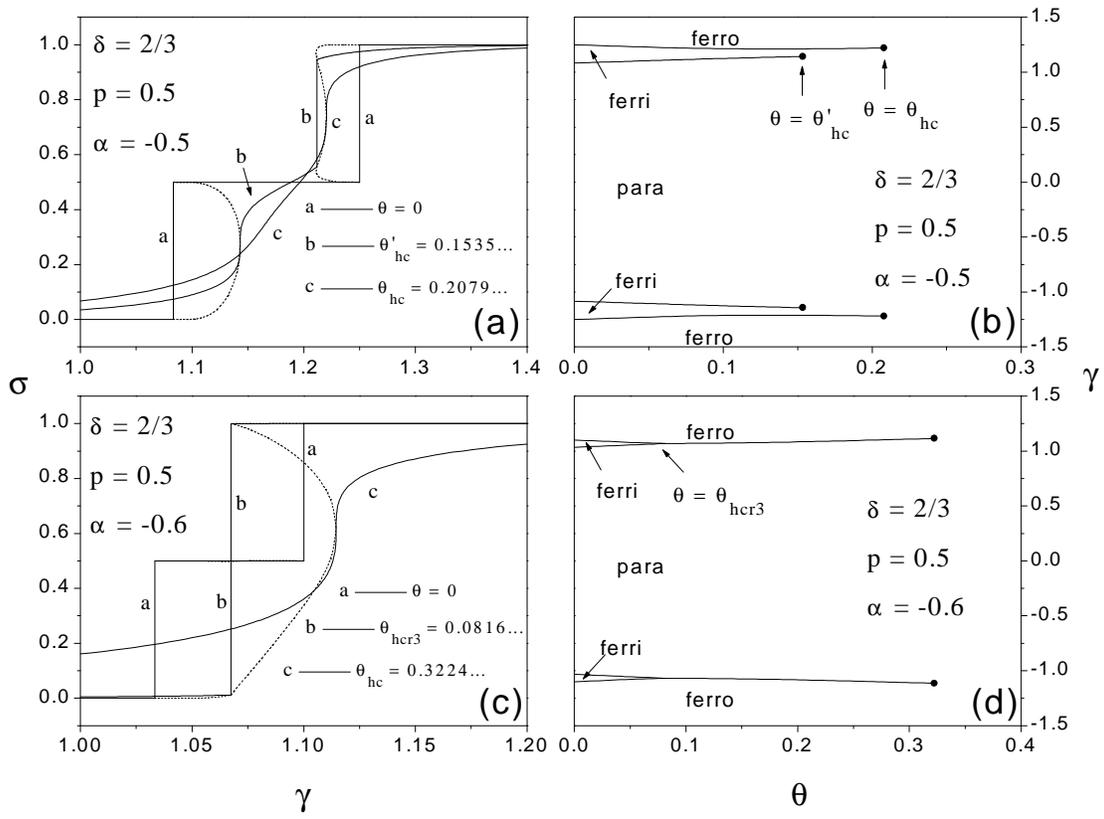
\noindent\textbf{Figure 1.} $\sigma\times\gamma(\gamma\equiv h/\left|
J_{B}\right|  )$ isotherms (a,c) and $\gamma\times\theta(\theta\equiv
k_{B}T/\left|  J_{B})\right|  $ phase diagrams (b,d) for $\delta=\frac{2}{3}$
($\delta\equiv J_{A}/J_{B}$) and $p=0.5$, with $\alpha=-0.5(\alpha\equiv
I/J_{B})$ (a,b) and $\alpha=-0.6$ (c,d). Dotted curves in (a) and (c)
correspond to phase coexistence boundaries, while full lines in (b) and (d)
indicate first order transitions.\bigskip

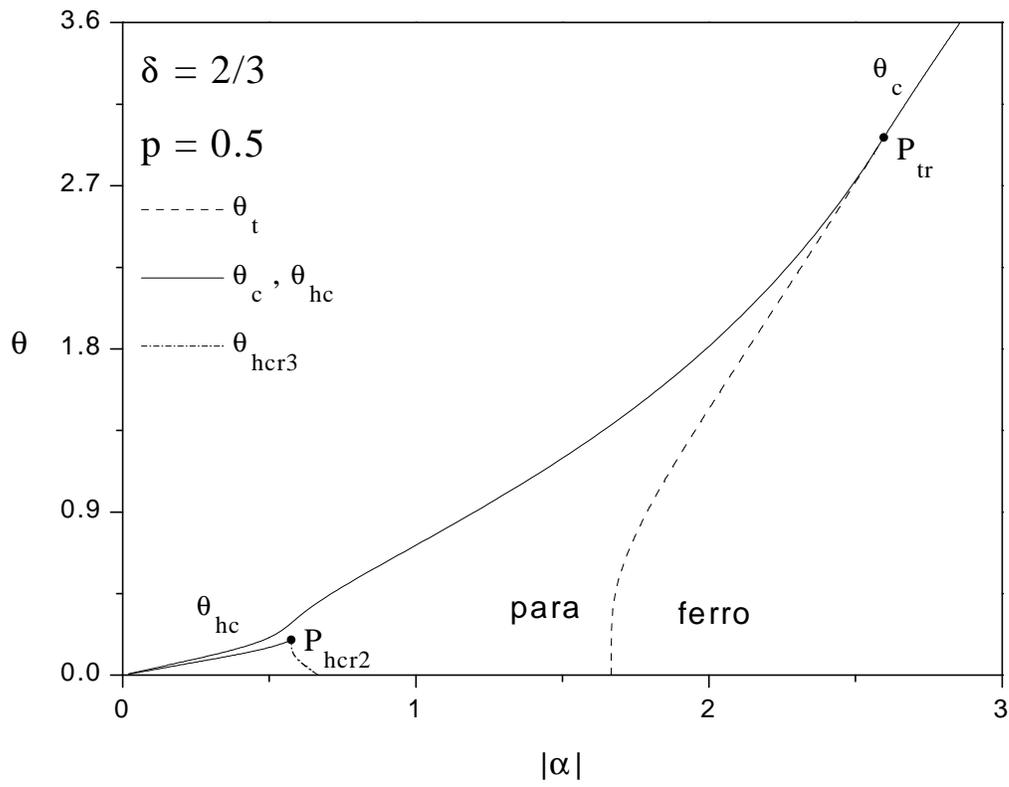
\noindent\textbf{Figure 2.} $\theta\times\alpha$ diagram for $\delta=\frac
{2}{3}$ and $p=0.5$. The dashed line shows the first order phase transition
temperature ($\theta_{t}$), which intersects the upper solid line
corresponding to second order transition temperatures ($\theta_{c}$ for $h=0$
and $\theta_{hc}$ for $h\ne0$) at the tricritical point $P_{tr}$. Temperature
$\theta_{hcr3}$ at which first order lines intersect in $\gamma\times\theta$
diagrams is shown in the dash-dotted curve. Point $P_{hcr2}$ marks the
intersection of that curve with another second-order curve. In this and all
the following $\theta\times\alpha$ diagrams labels ``para'' and ``ferro''
relate to regions in the diagram where at zero field the system is at
paramagnetic and ferromagnetic phases, respectively.\bigskip

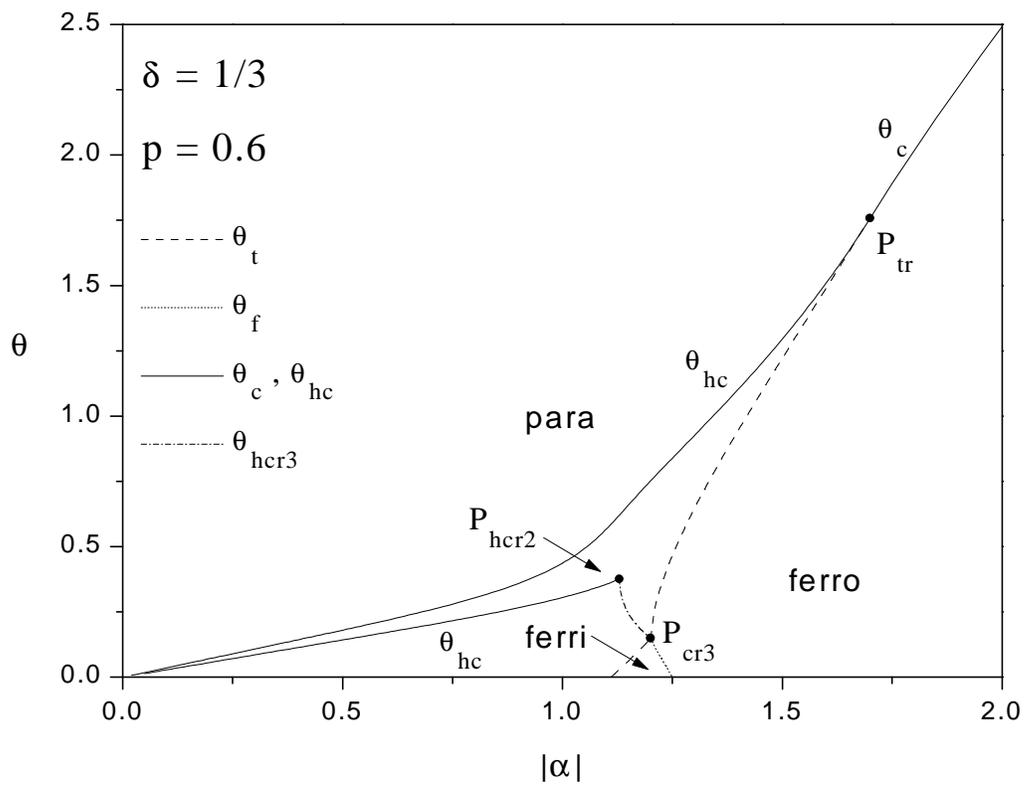
\noindent\textbf{Figure 3.} The same as in the previous figure for
$\delta=\frac{1}{3}$ and $p=0.6$. Solid and dashed curves represent second and
first-order paramagnetic transition temperatures, respectively. The dotted
curve shows the first-order ferri-ferromagnetic transition temperature
$\theta_{f}$. At point $P_{cr3}$, where $\theta_{hcr3}$, $\theta_{f}$ and two
$\theta_{t}$ curves intersect, paramagnetic, ferrimagnetic and ferromagnetic
phases coexist. The label ``ferri'' relates to the region where, at zero
field, the system is at the ferrimagnetic phase.\bigskip

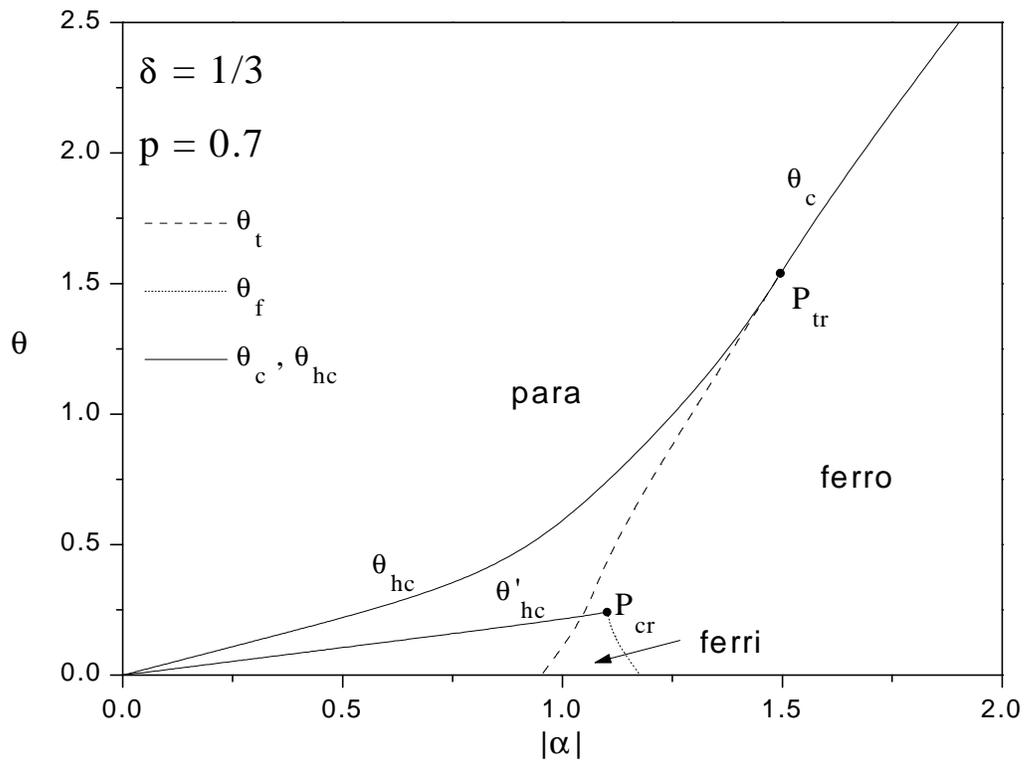
\noindent\textbf{Figure 4.} The same as in the previous figure for
$\delta=\frac{1}{3}$ and $p=0.7$. Differently from the previous diagram, there
is no three-phase coexistence point $P_{cr3}$ nor point $P_{hcr3}$, since the
$\theta_{f}$ curve ends at $P_{cr}$, reaching the $\theta_{hc}^{\prime}$ curve
before reaching the $\theta_{t}$ curve.\bigskip

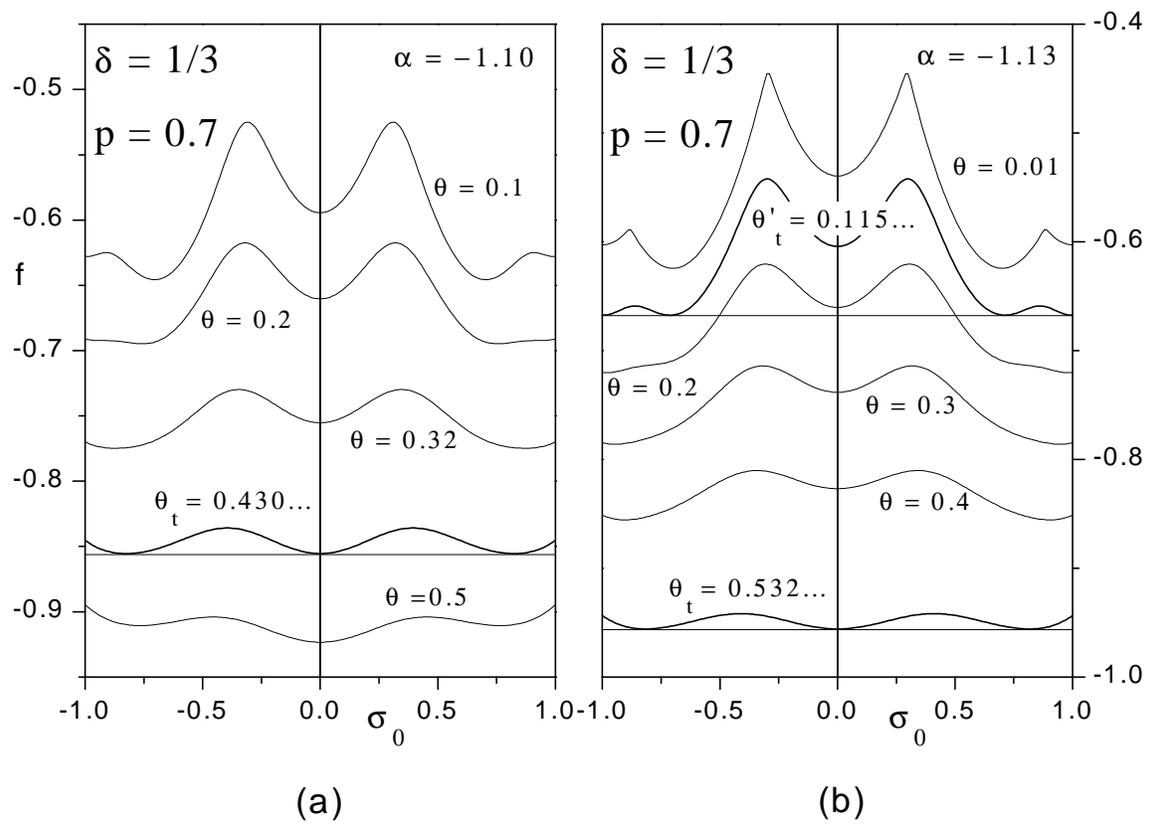
\noindent\textbf{Figure 5.} Renormalized free energy $f$ ($f\equiv
f_{a}/\left|  J_{B}\right|  $) as a function of the spontaneous magnetization
$\sigma_{0}\,$for various temperatures in the case $\delta=\frac{1}{3}$,
$p=0.7$ and $\alpha=-1.10$ (a) and $\alpha=-1.13$ (b). \bigskip

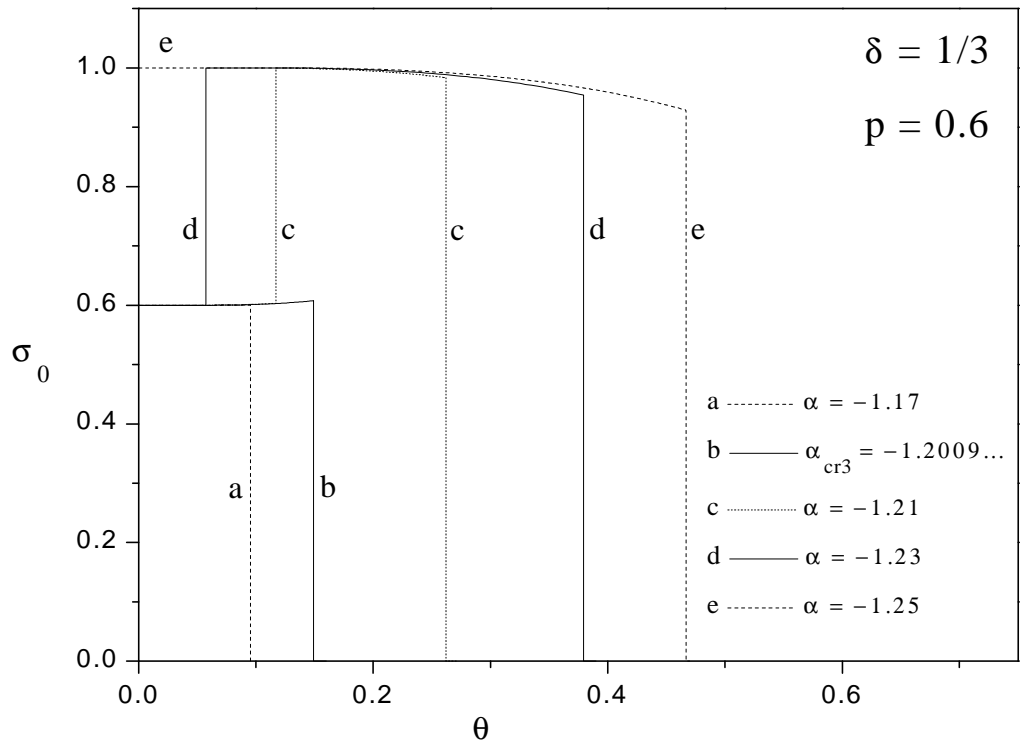
\noindent\textbf{Figure 6. }Spontaneous magnetization $\sigma_{0}$ as a
function of the renormalized temperature $\theta$ for $\delta=\frac{1}{3}$,
$p=0.6$ and various values of $\alpha$. \bigskip

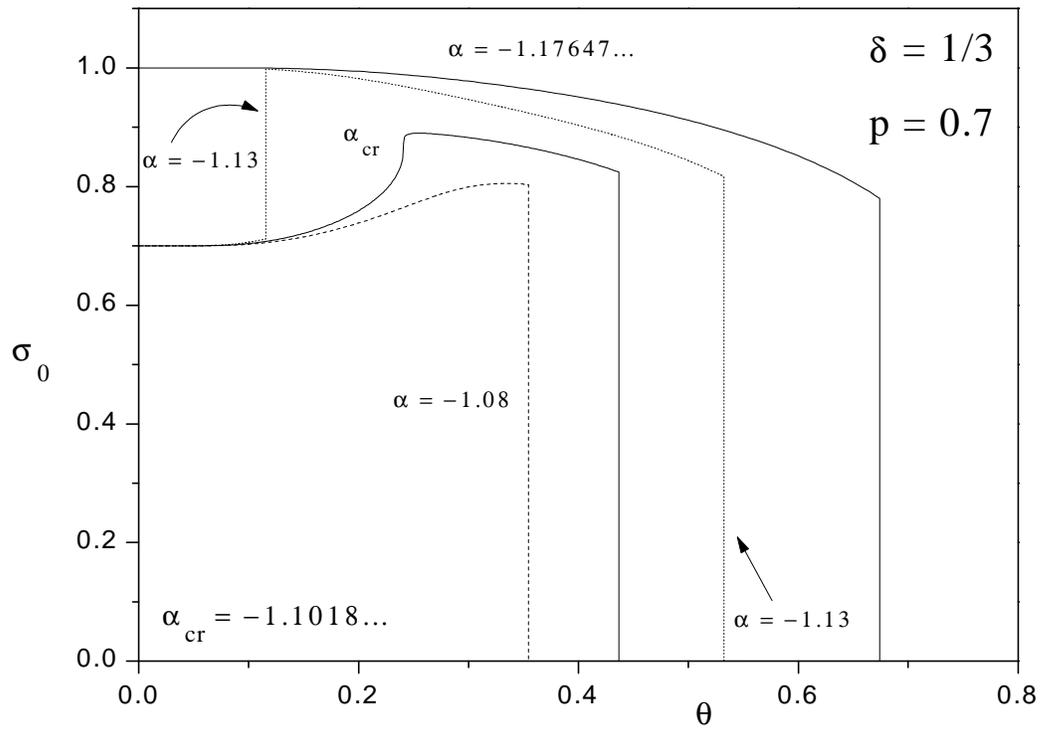
\noindent\textbf{Figure 7. }The same as in the previous figure for
$\delta=\frac{1}{3}$ and $p=0.7$. Notice that for $\alpha=\alpha_{cr}$ there
is a point in which $\sigma_{0}$ is a continuous function of the temperature
but the derivative diverges.\bigskip

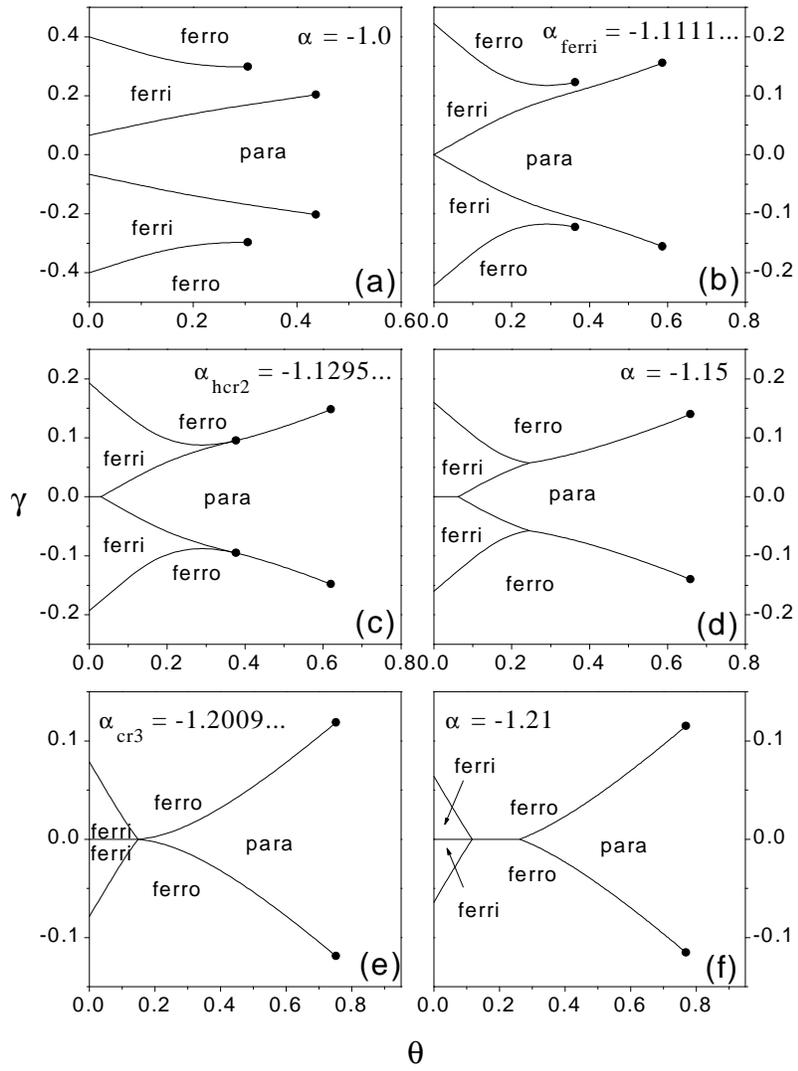
\noindent\textbf{Figure 8. }$\gamma\times\theta$ phase diagrams for
$\delta=\frac{1}{3}$, $p=0.6$ and values of $\alpha$ ranging from
$\alpha=-1\,$(a) to $\alpha=-1.21$ (f).\bigskip

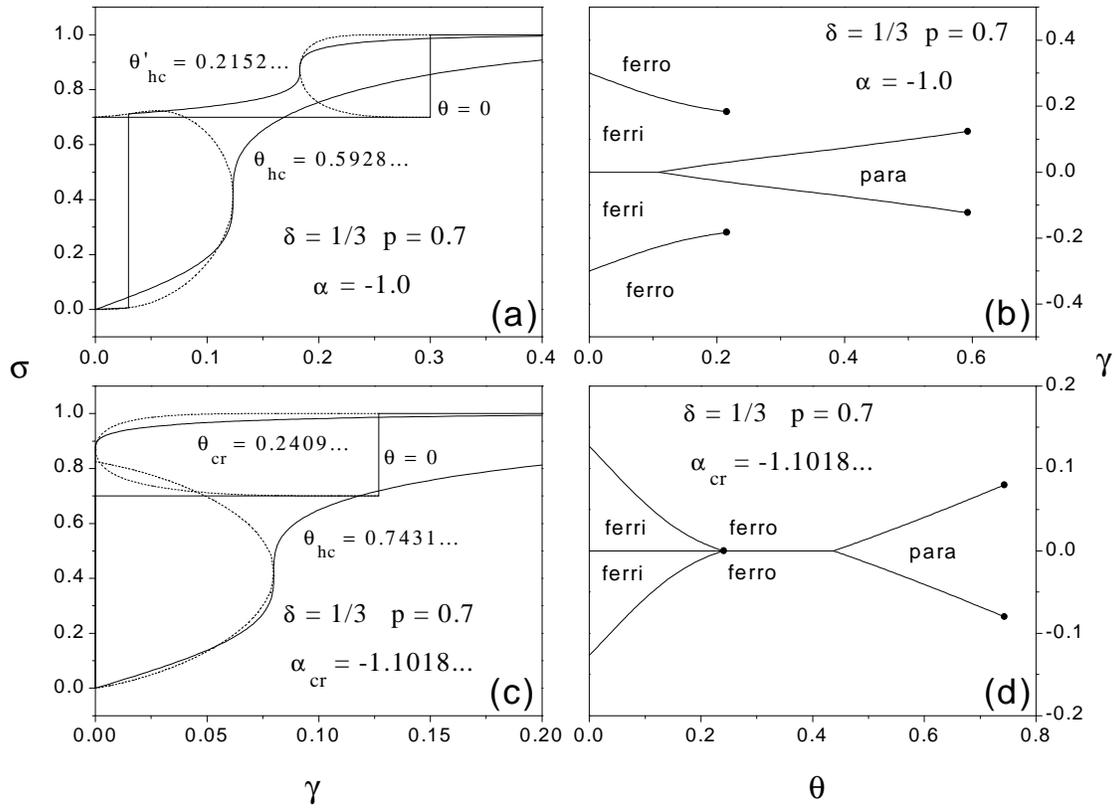
\noindent\textbf{Figure 9. }$\sigma\times\gamma$ isotherms (a,c) and
$\gamma\times\theta$ phase diagrams (b,d) for $\delta=\frac{1}{3}$ and
$p=0.7$, with $\alpha=-1$ (a,b) and $\alpha=\alpha_{cr}\simeq-1.1018$ (c,d). \bigskip

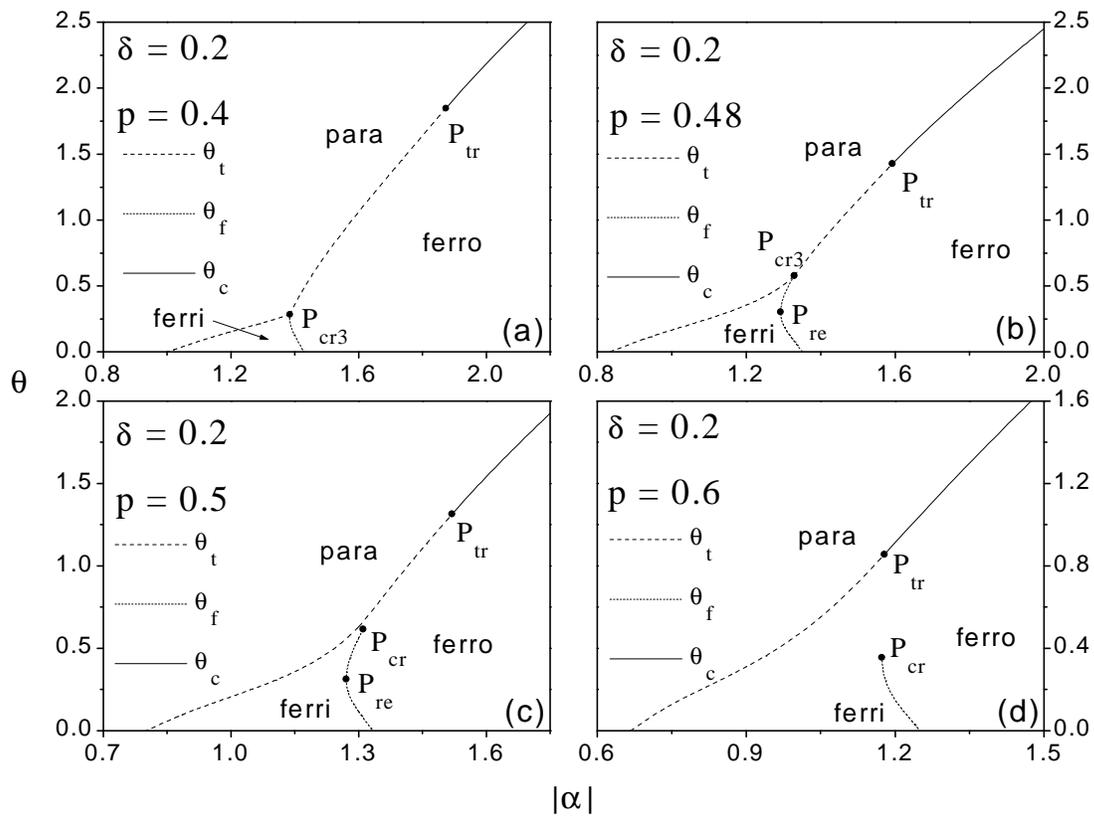
\noindent\textbf{Figure 10. }$\theta\times\alpha$ diagrams at zero field for
$\delta=\frac{1}{5}$ and $p=0.4$ (a), $p=0.48$ (b), $p=0.5$ (c) and $p=0.6$
(d). Notice the reentrant behaviour of the first-order ferri-ferromagnetic
transition lines, evident in (b) and (c).\bigskip

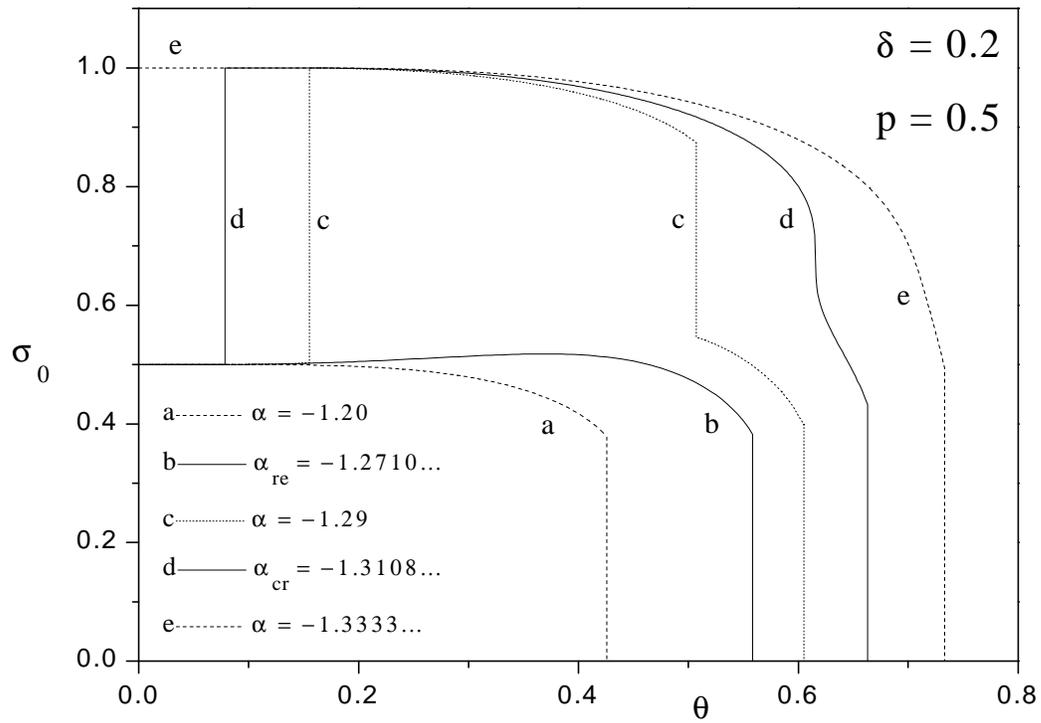
\noindent\textbf{Figure 11. }Spontaneous magnetization $\sigma_{0}$ as a
function of the renormalized temperature $\theta$ for $\delta=\frac{1}{5}$,
$p=0.5$ and various values of $\alpha$.\bigskip

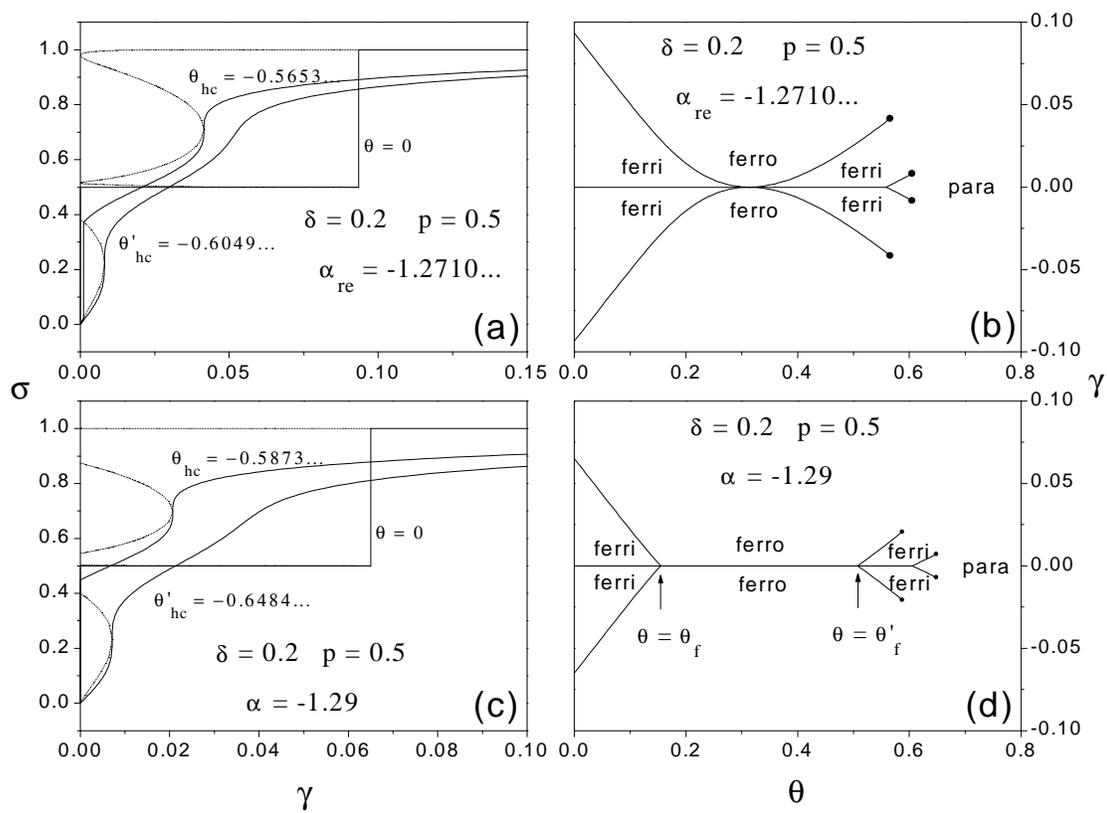
\noindent\textbf{Figure 12. }$\sigma\times\gamma$ isotherms (a,c) and
$\gamma\times\theta$ phase diagrams (b,d) for $\delta=\frac{1}{5}$ and
$p=0.5$, with $\alpha=\alpha_{re}\simeq-1.2710$ (a,b) and $\alpha=-1.29$ (c,d).\bigskip

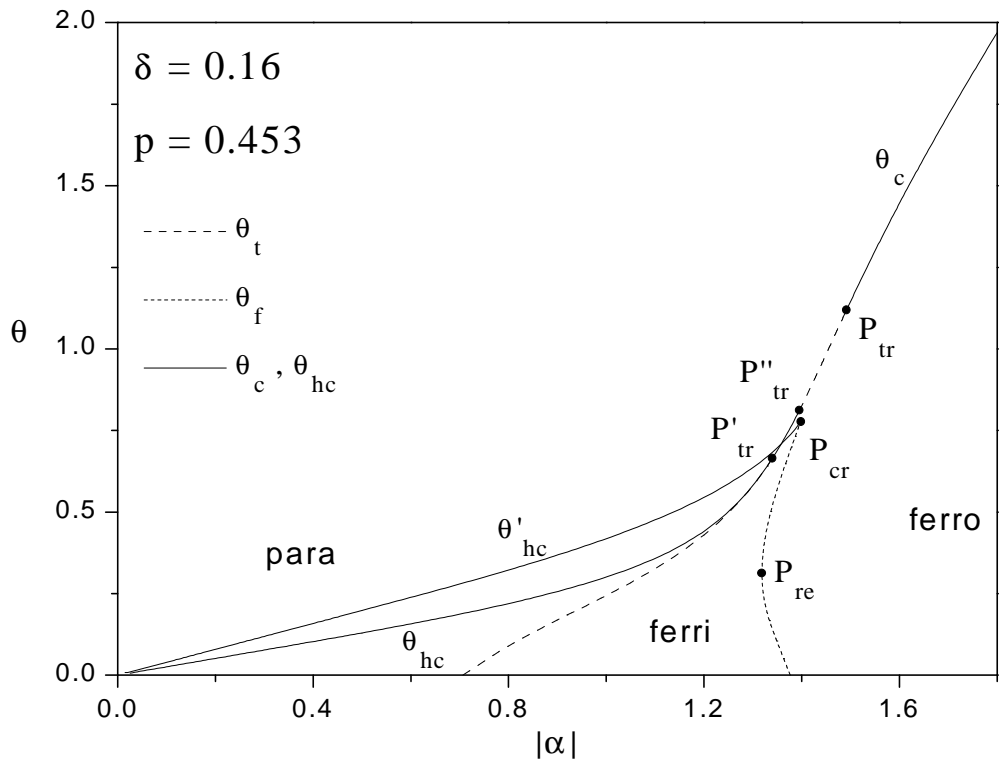
\noindent\textbf{Figure 13. }$\theta\times\alpha$ diagram for $\delta=0.16$
and $p=0.453$. Notice the existence of three tricritical points ($P_{tr}$,
$P_{tr}^{\prime}$ and $P_{tr}^{\prime\prime}$) and of another zero-field
second-order paramagnetic transition for $\alpha_{tr}^{\prime\prime}%
<\alpha<\alpha_{tr}^{\prime}$. For $\alpha_{tr}^{\prime}<\alpha<\alpha_{tr}$
there is a third $\theta_{hc}$ line just above the $\theta_{t}$ line, but
omitted from the figure for clarity.\bigskip

\noindent\textbf{Figure 14. }$\theta\times\alpha$ diagram for $\delta=0.15$
and $p=0.42$. The first-order ferri-ferromagnetic transition line ($\theta
_{f}$) intersects the second-order ferri-paramagnetic transition line
($\theta_{c}^{\prime}$) and the first-order ferro-paramagnetic transition line
($\theta_{t}$) at point $P_{c2}$, which is \textit{not} a tricritical point.\bigskip

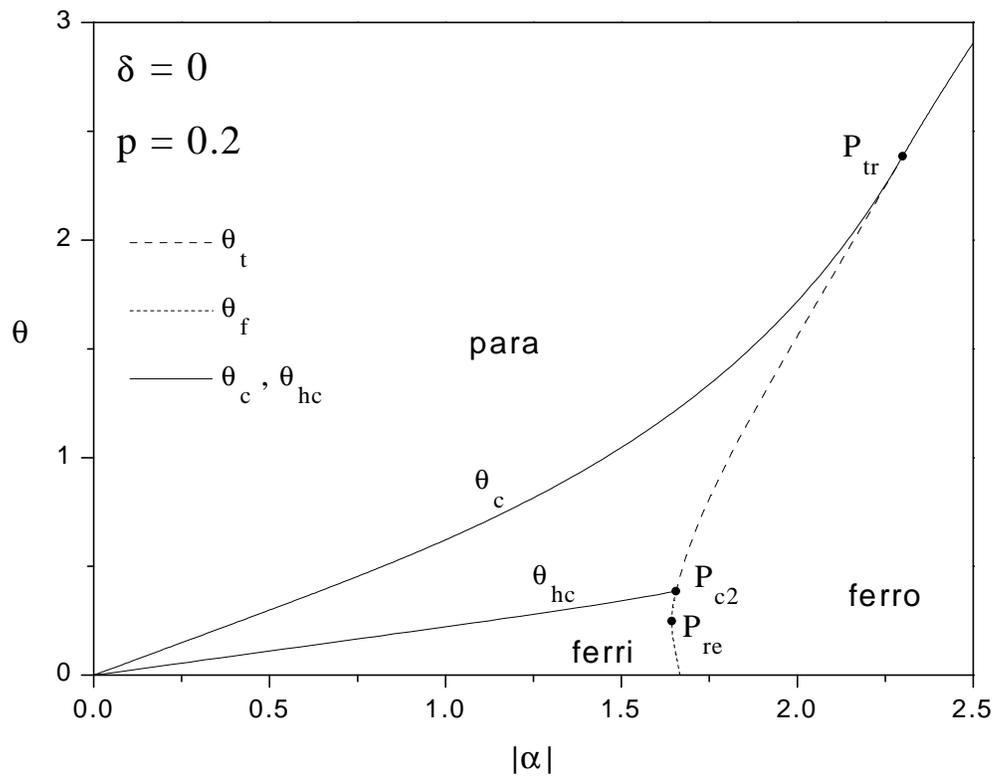
\noindent\textbf{Figure 15. }$\theta\times\alpha$ diagram for $\delta=0$ and
$p=0.2$.\bigskip

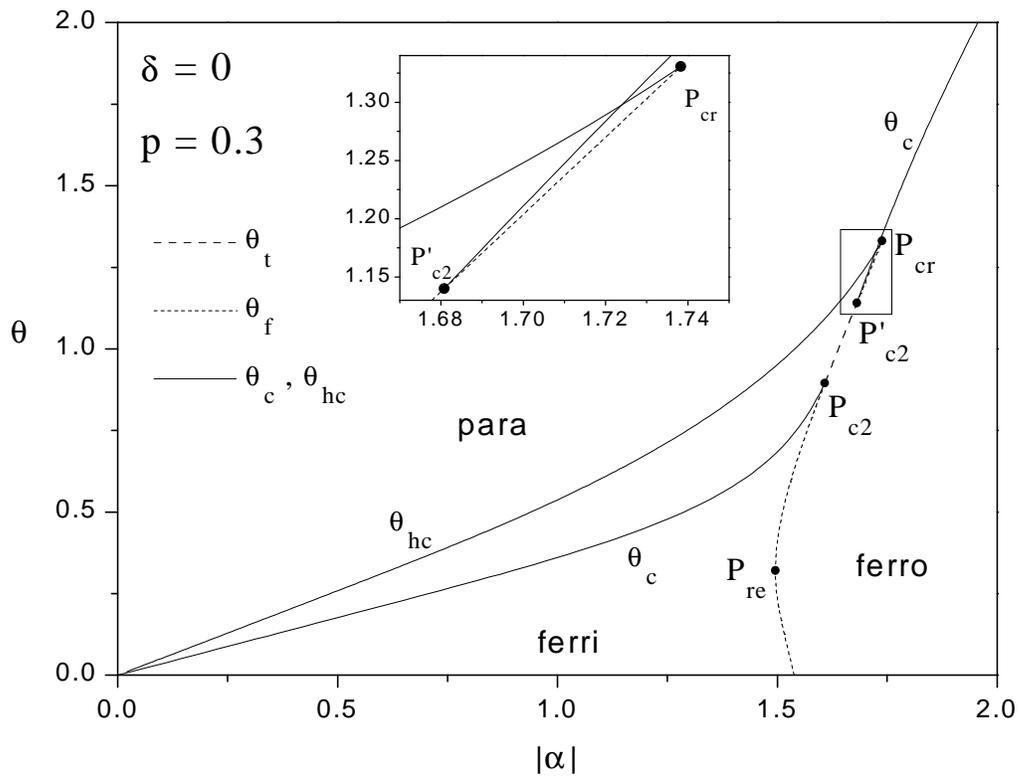
\noindent\textbf{Figure 16. }$\theta\times\alpha$ diagram for $\delta=0$ and
$p=0.3$. The dotted line in the inset corresponds to fist-order
ferri-ferromagnetic transition temperatures ($\theta_{f}$).\bigskip

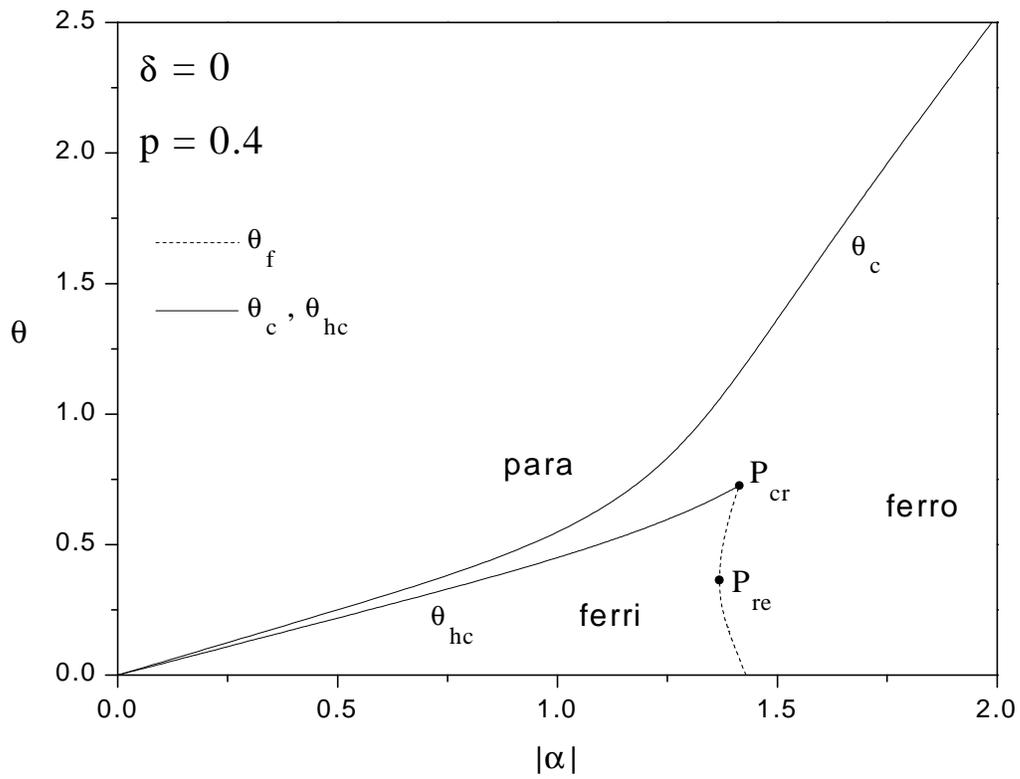
\noindent\textbf{Figure 17. }$\theta\times\alpha$ diagram for $\delta=0$ and
$p=0.4$.\bigskip

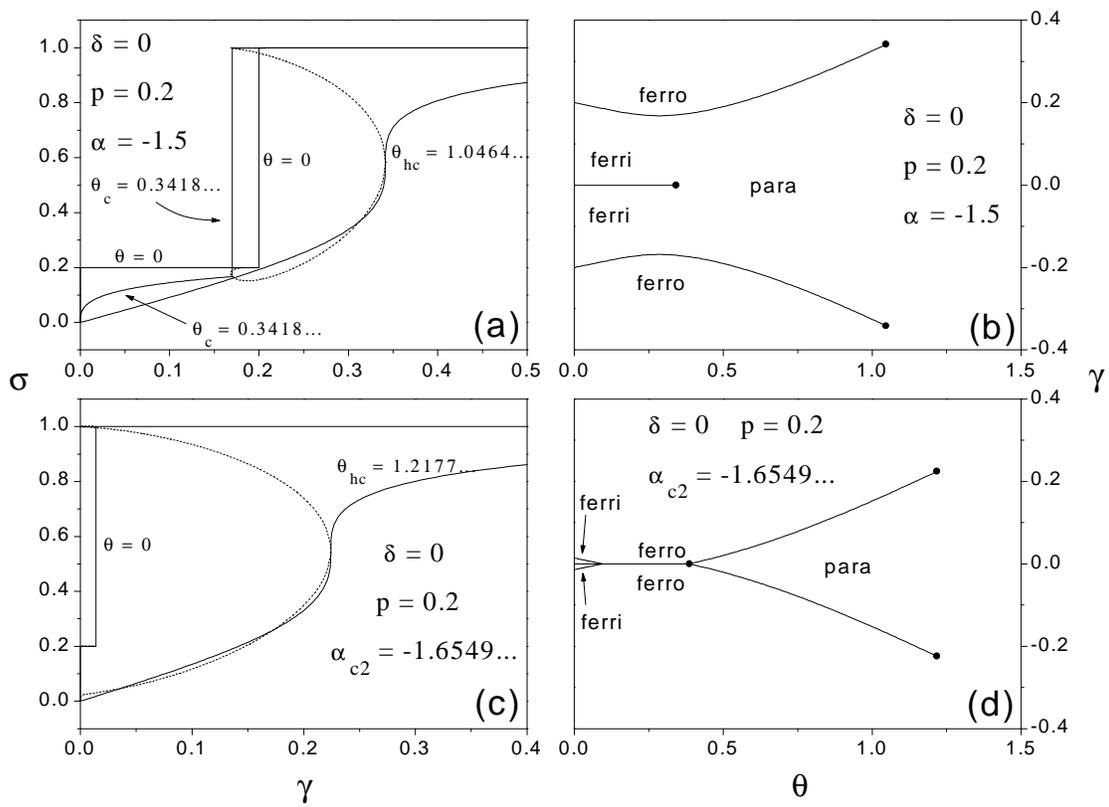
\noindent\textbf{Figure 18. }$\sigma\times\gamma$ isotherms (a,c) and
$\gamma\times\theta$ phase diagrams (b,d) for $\delta=0$ and $p=0.2$, with
$\alpha=-1.5$ (a,b) and $\alpha=\alpha_{c2}\simeq-1.6549$ (c,d).\bigskip

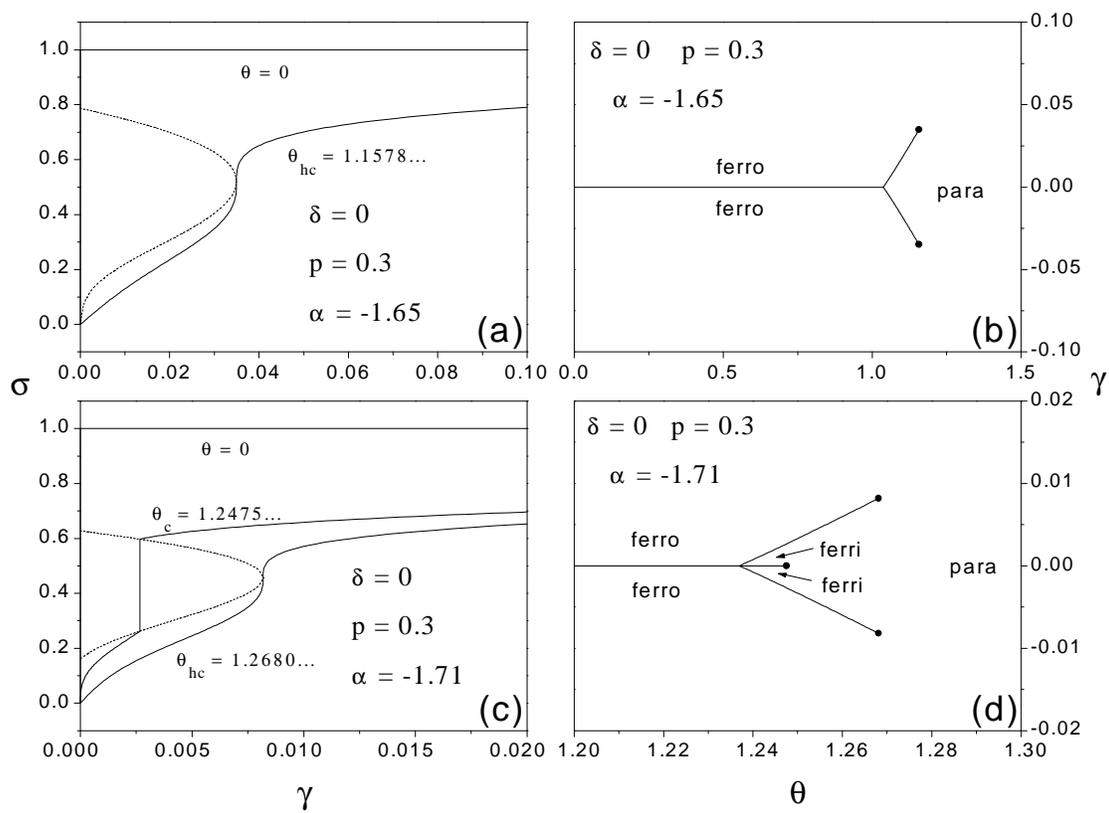
\noindent\textbf{Figure 19. }$\sigma\times\gamma$ isotherms (a,c) and
$\gamma\times\theta$ phase diagrams (b,d) for $\delta=0$ and $p=0.2$, with
$\alpha=-1.65$ (a,b) and $\alpha=-1.71$ (c,d).
\end{document}